# Detection of Lens Candidates for the Double QSO Q2345+007

Philippe Fischer [1], J. Anthony Tyson [1] [2] [3], Gary M. Bernstein[2] [3] [4], and Puragra Guhathakurta[2] [3] [5]

astro-ph/9405062  26 May 94

[1]AT&T Bell Laboratories, 600 Mountain Ave., Murray Hill, NJ 07974

[2]Visiting Astronomer, Canada France Hawaii Telescope

[3]Visiting Astronomer, National Optical Astronomy Observatories, which is operated by the Association of Universities for Research in Astronomy, Inc., under contract to the National Science Foundation.

[4]Steward Observatory, University of Arizona, Tucson, AZ 85721

[5]Space Telescope Science Institute, 3700 San Martin Dr., Baltimore, MD 21218

— 2 —

## ABSTRACT

Luminous objects associated with the 7.06″ separation double quasar Q2345+007 have, until now, escaped detection. In this letter we present the results of the deepest known imaging of the region surrounding the quasar. The total exposure times were 47600 seconds in $B_j$ (101 frames) and 32160 seconds in R (82 frames). The frames came from CFHT, CTIO, and KPNO. We detect a $B_j = 25.0$ mag galaxy ($B_j - R = 0.5$ mag) in close proximity to the fainter QSO image, possibly at $z = 1.49$ given by several absorption features in the QSO spectra. Furthermore, there is a $2.9\sigma$ enhancement in the number density of faint galaxies ($24 \leq B_j \leq 28$, $B_j - R \leq 1.5$) near the quasar and another $3.1\sigma$ enhancement further away. These discoveries support the hypothesis that Q2345+007 is being lensed by one or more distant mass concentrations and may imply that compact "clusters" already exist at early epochs. We discuss several lens models for the system and the cosmological implications.

*Subject headings:* dark matter - galaxies: clustering - gravitational lensing - quasars: individual(Q2345+007)

## 1. Introduction

The evolution of cluster-size concentrations and their expected numbers and peak densities at moderate redshifts are related to the nature of the dark matter and the ratio of its cosmic density to the critical density, $\Omega$. The existence of compact mass concentrations of $10^{13} - 10^{14}$ $M_\odot$ at large redshift would have significant implications for cosmology and may favor a low value of $\Omega$ (Cen, Ostriker & Peebles, 1993).

In this context, Q2345+007AB has attracted considerable attention. More than a decade after its first discovery (Weedman *et al.* 1982) there is still controversy regarding its true nature. Is this a lensed system or simply two quasars with remarkably similar spectra and redshifts (z = 2.15, $\Delta V$ = 15 $\pm$20 km s$^{-1}$; Steidel & Sargent 1991)? Despite having the largest separation of any known double quasar, no good lens candidate has been found in deep broadband images (Tyson *et al.* 1986).

Both the QSO images have metal absorption lines with redshift z = 1.491 (Foltz et al 1984, Steidel & Sargent 1991) and the B image has a second C IV doublet ($z = 1.483$) with $\Delta V = 1060 \pm 60$ km s$^{-1}$. Is this evidence for lensing by a large cluster at $z = 1.49$? Absorption lines at other redshifts in the range $z = 0.75 - 1.98$ are also seen although the largest number (15) are from the two systems at $z = 1.483$ and $z = 1.489$. Further evidence comes from Bonnet et al. (1994) who claim the detection of a possible lens from weak lensing arclet candidates near the QSO.

Here we report on the discovery of candidate lenses projected near Q2345+007. In §2. we describe the detection of a galaxy near the B QSO image and two significant galaxy number density enhancements. In §3. we present HST astrometry of the quasars, and investigate previous claims for the existence of subcomponents. In §4. we describe some lens models which fit the observations. The main conclusions are summarized in §5.

## 2.   Is Q2345+007 A Lensed Quasar?

We have constructed very deep $B_j$ and R images from the nine different epochs listed in Table 1. The total exposure times were 47600 seconds in $B_j$ (101 frames) and 32160 seconds in R (82 frames). Since the seeing and signal-to-noise of the individual frames varied we used an optimal weighting algorithm to combine them (Fischer & Kochanski 1994). The RMS scatter in the sky backgrounds are approximately 31.2 mag arcsecond$^{-2}$ ($B_j$) and 29.9 mag arcsecond$^{-2}$ (R). It should be noted, however, that the noise on our final frames is correlated (non-white). Fig. 1 is a color image of the region.

### 2.1.   Detection of a Candidate Lens Galaxy

Since there are strong metal absorption lines in the QSO spectra, it is likely that there are galaxies projected close to the QSO images. Ideally, one would like to model the stellar point spread function (PSF) and subtract the QSO images. The complete overlap region of all the frames is approximately $80'' \times 80''$ and contains no bright stars; it is impossible to obtain a useful PSF independent of the QSO. We, instead, subtracted a scaled version of A from B, and B from A. The scaling and centering were determined from iterative weighted least-squares fitting The A/B flux ratio is approximately 3.4 in $B_j$ and 3.1 in R. Lensing should be achromatic; differences can arise due to: variability of the QSO and the net timing delay between A and B, or differential reddening. Alternatively, this may be evidence against the lensing hypothesis. This issue could be settled with long-term photometric monitoring. The mean brightnesses for the 9 individual epochs of image A are $B_j = 19.33 \pm 0.08$ mag, R = $18.78 \pm 0.05$ mag, while image B has $B_j = 20.64 \pm 0.06$ mag and R = $19.96 \pm 0.03$ mag.

In both the subtracted $B_j$ and R images (Fig 2) there is a source (G1) located 1.0 - 1.2″ north and 0.8″ east of image B (0.4 - 0.6″ northwest of the line joining A to B, see Table 2). This detection is highly significant; the peak intensity of G1 is greater than one third the subtracted $B_j$ flux at that position, and one sixth the R flux. In both cases, the peaks are greater than twelve sky standard deviations. The centers agree at the $0.2''$ level between the $B_j$ and R frames. The brightness of G1 is $B_j = 25.0 \pm 0.5$ and $R = 24.5 \pm 0.5$. These convert to: B(Johnson) = 25.0, V(Johnson) = 24.6, $R_c$(Cousins) = 24.4 (Gullixson *et al.* 1994).

## 2.2. Detection of a Candidate Lens Cluster

We have run FOCAS on QSO-subtracted versions of the $B_j$ and R images. Fig 3 shows number density contours for galaxies with $24 \leq B_j \leq 28$ and $B_j - R \leq 1.5$ The galaxy number density map was determined using an adaptive kernel estimator (Silverman 1986). There is a density enhancement located $17''$ north and $0.7''$ west of G1. There are two other large maxima located further to the west. The detection of the large density maximum near the QSO images is very robust to changes in how the galaxies were culled from the catalogs (e.g. relatively independent of color and brightness cut-offs), however, the position is somewhat less robust; the uncertainties are several arcseconds. If we are missing faint galaxies projected near the QSO images, this would tend to bias the center away from the quasar. The detection nearest the QSO is significant at around the $2.9\sigma$ level; a $40'' \times 40''$ box centered on the density enhancement contains 49 galaxies while 32.4 are expected. The enhancement just to the west is significant at the $3.1\sigma$ level (50 galaxies). These density enhancements may be one or more clusters at large redshift.

Because not all regions of this field have the same coverage, we have carried out artificial galaxy simulations; galaxies were added to the frame, FOCAS was run, and completeness histograms were constructed In the region shown in Fig. 3, there are no substantial variations in the completeness fractions as a function of position for the detection thresholds we have employed.

## 3. HST Images

Four 900s exposures of Q2345+007 were obtained with the Hubble Space Telescope wide field planetary camera on 1991 May 15, two with the F555W filter and two with the F785LP filter. Image positions were measured using the STSDAS task METRIC; the image separation is $7.06 \pm 0.01''$. There have been two previous claims that component B has an extended appearance perpendicular to the line to A (Nieto *et al.* 1988, Weir & Djorgovski 1991). The former also found that B was split into two similar intensity images, B1 and B2 aligned pointing towards A. The existence of subcomponents would have important implications for the lensing hypothesis, however, in the combined HST image, there is no evidence for elongation or splitting in image A or B at the brightness levels previously claimed.

Using Tiny Tim (Krist 1992), we generated artificial PSFs at the relevant positions and bandpasses. These had FWHM of approximately 1.6 pixels for F555W and 2.1 pixels for F785LP, smaller than the observed values ($3.4''$- $3.8''$). About $0.045''$ of spacecraft "jitter" is required to produce this broadening, at the high end of the normal range during quiescent periods ($0.015''$ to $0.050''$, Burrows *et al.* 1991), and we, therefore, put an upper limit to the quasar FWHM of $\leq 0.10''$. At this level we contradict both of the previous claims, and conclude that their results were artifacts of the image restoration techniques employed.

## 4. Lens Models

In this section we assume that Q2345+007 is a lensed system and show that there are some simple lens models that provide good fits to the data. At present there are insufficient constraints to uniquely determine a lensing solution. The known parameters are: the image positions, the angular positions of G1 and the cluster (C1), and the flux ratio of the QSO images (3.25). The flux ratio is unreliable since the QSO image brightnesses are varying and there is always the possibility of microlensing. The *assumed* redshift of the lens(es) is/are $z_d = 1.49$ and the *known* QSO redshift is $z_s = 2.15$. All the lens models we will consider assume thin transparent lenses, an Einstein-de Sitter universe, and $H_0 = 75 \text{ km s}^{-1} \text{ Mpc}^{-1}$.

To produce a splitting of 7.06″, a single point-mass would have to be approximately $10^{13}$ $M_\odot$ if it were located 1.5″ from one of the images and about $1.5 \times 10^{13}$ if it were equidistant from both. An extended object would require an even higher total mass. A single, circularly symmetric, object located at G1 could not produce the observed "dogleg". This requires non-circular and/or multiple lenses.

Following Kormann *et al.* (1994), we construct "nonsingular isothermal ellipsoid" lenses with surface densities of the form

$$\Sigma(R) = \frac{\sigma^2}{2G} \frac{\sqrt{f}}{\sqrt{R^2 + R_c^2}}, \tag{1}$$

where $\sigma$ is the velocity dispersion, $f$ is the axis ratio, $R = x^2 + f^2 y^2$, and $R_c$ is the angular core radius. The relationship between the source position $(u, v)$ and the image position $(x, y)$ is given in Kochanek (1991) along with expressions for image magnification.

Model 1 consists of a single lens positioned at G1, and, as shown in Table 2 provides an excellent fit to the observables. This model has $\sigma = 850$ km s$^{-1}$. As a point of reference, the giant Virgo elliptical M87 has $\sigma = 384 \pm 42$ km s$^{-1}$ and has kinematics consistent with an isothermal sphere with $R_c = 5.81$ kpc (Merritt & Tremblay 1993). Our best fit model with this core radius has $\sigma > 1000$ km s$^{-1}$ for G1. Therefore, it is unlikely that a single massive galaxy at the position of G1 is solely responsible for the observed lensing. However, a cluster centered at G1 with a core radius of 100 kpc and a velocity dispersion of $\sim 1700$ km s$^{-1}$ would reproduce the observations well. This is similar to the velocity dispersion of the large cluster Abell 1689 ($1850 \pm 170$ km s$^{-1}$, Gudehus 1989, although the lens solution yields $1350 \pm 75$ km s$^{-1}$ Tyson et al. 1990).

Model 2 assumes two lenses, G1 and the closest candidate cluster, C1. Once again the observables can be well fit as shown in Table 2. Model 2 has a very high ellipticity and $\sigma$. It is possible to reduce this ellipticity while maintaining a good fit, but this requires an even higher $\sigma$. This model represents roughly the maximum $\sigma$ one might expect from a galaxy (400 km s$^{-1}$). As mentioned above, the exact position of C1 is quite uncertain, and might in fact be closer to G1. If we arbitrarily decrease the angular distance in the northerly direction from $17.22''$ to $8.61''$ while maintaining the same ellipticity we find that the velocity dispersion decreases from 2840 km s$^{-1}$ to 1840 km s$^{-1}$, and the core radius to $8.4''$.

There is further information available regarding the field surrounding Q2345+007. Redshifts have been measured for three of the brighter red galaxies, yielding z = 0.279 - 0.280 (Bonnet et al. 1994), evidence for the presence of a foreground cluster. For Model 3 we have included this cluster, treating it as a uniform sheet with surface density $\Sigma_{C2}$. The equations for lensing by mass distributions at two redshifts are given in Kochanek & Apostolakis (1988).

The parameters for Model 3 are shown in Table 2. By introducing a sheet with a surface density of $\Sigma_{C2} = 0.5$ g cm$^{-2}$, corresponding to a cluster with a velocity dispersion of about 800 km s$^{-1}$ and a core radius of 90 kpc ($23''$ for $\Omega = 0$), we were able to reduce the velocity dispersion of C1 to 1000 km s$^{-1}$, consistent with the splitting seen in the CIV lines.

Large density enhancements also cause distortions in the appearance of faint blue background galaxies (FBGs) (Tyson et al. 1990), although most of the FBGs probably lie at redshifts less than 1.5. Bonnet, et al. (1994) claim to detect some weak lensing candidates near Q2345+007. However, their result is difficult to asses since they do not include information on the brightness, color or number of arclet candidates. We will discuss weak lensing in our data for Q2345+007 in a separate paper.

At $z = 1.5$, the I band roughly corresponds to the rest-frame B$_j$ band. The faint galaxies in this field have B$_j$–R$\sim$ 0–1 for $26 \leq$B$_j \leq 27$, consistent with the FBG population found in other surveys (Tyson 1988). Assuming that G1 and the other two faint galaxies near the QSO circle have R–I$\sim$ 1 like the FBGs, their summed I = 23 mag converts to an absolute magnitude of −22.3 to −23 mag and a blue luminosity of 1.5–2.8 × $10^{11}$ $L_{B\odot}$ (1Ω0) at z = 1.5. This implies an absorption-free M/$L_B$ = 60–80 M/$L_{B\odot}$ inside the circle bounded by the quasar images.

## 5. Conclusion

In this paper we have endeavored to find lens candidates for the double QSO 2345+007. We constructed R and $B_j$ images from exposures taken at CFHT, CTIO, KPNO; the total exposure times were 47600s in $B_j$ (101 frames) and 32160 in R (82 frames).

The conclusions are: 1) there is a faint galaxy ($B_j$ = 25.0 mag, $B_j$ – R = 0.5 mag) located 1.5$''$ from the B image. 2) There is a significant excess of galaxies with 24 ≤ $B_j$ ≤ 28 and $B_j$ – R ≤ 1.5 near the QSO, evidence for clustering at large redshift. 3) There is no evidence for subcomponents at the 0.1$''$ level, contradicting two previous claims. 4) It is possible to reproduce the image positions and flux ratios using simple lens models although the problem is currently underconstrained. Multiple lenses with a foreground mass screen are favored. 5) The evidence we have uncovered in this letter supports the lensing hypothesis. To conclusively decide the issue will require long-term photometric monitoring of the QSO images to search for a timing delay.

Gravitational lensing directly measures M($\leq$R)/R. If the lensing is occurring at z = 1.5, then a mass of at least M(R$\leq$ 20kpc)$\sim$ $10^{13}$ $M_\odot$ is required. There is evidence for a large lens at $z$ = 1.49: the previously detected absorption line systems separated by 1000 km s$^{-1}$ and the overdensity of faint galaxies reported in this letter. If future redshift determinations confirm the cluster at $z$ = 1.49, then our result has direct cosmological implications. The existence of massive compact structures as early as z=1.5 is less likely in cold, hot or mixed dark matter models ($\Omega = 1$) than in an open ($\Omega = 0.2$) primordial baryonic isocurvature model (see Cen, Ostriker & Peebles 1993, Fig. 13) which can form structure earlier.

PF thanks NSERC and Bell Labs for postdoctoral fellowships. Thanks to P. Boeshaar, R. Cen, C. Christian, K. Gebhardt, P. Seitzer, P. Waddell, D. Weinberg and R. Wenk.

---

Table 1: Journal of Observations

| Date (dd/mm/yy/) | Observatory | CCD | Scale ($''$/pix) | Filter | Exp. | FWHM ($''$) |
|---|---|---|---|---|---|---|
| 24/10/84 - 25/10/84 | CTIO | RCA $320 \times 508$ | 0.59 | $B_j$ | $7 \times 900$s | 1.6 - 1.9 |
| 23/10/84 | CTIO | RCA $320 \times 508$ | 0.59 | $B_j$ | $1 \times 500$s | 1.5 |
| 23/10/84 | CTIO | RCA $320 \times 508$ | 0.59 | R | $11 \times 500$s | 1.2 - 1.3 |
| 23/10/84 - 24/10/84 | CTIO | RCA $320 \times 508$ | 0.59 | I | $14 \times 500$s | 1.1 - 1.4 |
| 09/11/85 - 10/11/85 | CTIO | RCA $320 \times 508$ | 0.59 | B | $8 \times 500$s | 1.3 - 1.6 |
| 17/09/87 - 19/09/87 | CFHT | RCA $340 \times 528$ | 0.41 | $B_j$ | $31 \times 500$s | 0.8 - 1.1 |
| 17/09/87 - 19/09/87 | CFHT | RCA $340 \times 528$ | 0.41 | R | $11 \times 500$s | 0.9 - 1.1 |
| 24/10/89 - 25/10/89 | CFHT | RCA $340 \times 512$ | 0.41 | $B_j$ | $9 \times 500$s | 0.8 - 1.0 |
| 24/10/89 - 25/10/89 | CFHT | RCA $340 \times 512$ | 0.41 | R | $9 \times 500$s | 0.7 - 1.0 |
| 18/08/90 | KPNO | Tek $1024 \times 1024$ | 0.47 | $B_j$ | $3 \times 300$s | 1.2 |
| 17/08/90 | KPNO | Tek $1024 \times 1024$ | 0.47 | R | $3 \times 300$s | 1.0 - 1.1 |
| 23/09/90 - 24/09/90 | CTIO | TI $800 \times 800^*$ | 0.58 | $B_j$ | $20 \times 300$s | 1.2 - 1.5 |
| 23/09/90 - 24/09/90 | CTIO | TI $800 \times 800^*$ | 0.58 | R | $25 \times 300$s | 1.1 - 1.3 |
| 08/09/91 | KPNO | Tek $1024 \times 1024$ | 0.47 | $B_j$ | $3 \times 300$s | 1.2 - 1.3 |
| 08/09/91 | KPNO | Tek $1024 \times 1024$ | 0.47 | R | $3 \times 300$s | 1.0 - 1.1 |
| 08/12/91 | KPNO | Tek $2048 \times 2048$ | 0.47 | R | $3 \times 720$s | 1.3 - 1.4 |
| 26/10/92 - 27/10/92 | CTIO | Tek $1024 \times 1024$ | 0.48 | $B_j$ | $12 \times 500$s | 1.3 - 1.7 |
| 26/10/92 - 27/10/92 | CTIO | Tek $1024 \times 1024$ | 0.48 | R | $13 \times 400$s | 1.2 - 1.6 |
| 23/06/93 | CTIO | Tek $2048 \times 2048$ | 0.48 | $B_j$ | $6 \times 500$s | 1.2 - 1.5 |

$^*$Rebinned to $400 \times 400$

Table 2: Lens Models

| | Observed | Model 1 | | Model 2 | Model 3 |
|---|---|---|---|---|---|
| | | $(\Omega = 0)$ | $(\Omega = 1)$ | $(\Omega = 0)$ | $(\Omega = 0)$ |
| $(x,y)_{G1}$ ($''$) | 0.00, 0.00 | | | | |
| $(x,y)_A$ ($''$) | 5.08, 2.90 | 5.08, 2.90 | 5.08, 2.90 | 5.08,–2.90 | 5.08, 2.90 |
| $(x,y)_B$ ($''$) | –0.78,–1.04 | –0.78,–1.04 | –0.78,–1.04 | –0.78,–1.04 | –0.78,–1.04 |
| $(x,y)_{C1}$ ($''$) | –0.72,17.22 | $\cdots$ | $\cdots$ | –0.72,17.22 | –0.72,17.22 |
| $(x,y)_{QSO}$ ($''$) | $\cdots$ | 1.73, 1.41 | 1.73, 1.41 | 3.72, 8.05 | 0.20, 0.73 |
| Magnification (A,B) | $\cdots$ | 1.96,–0.60 | 1.96,–0.60 | 7.80, 2.40 | 30.54,–9.39 |
| Flux Ratio (A/B) | 3.25 | 3.25 | 3.25 | 3.25 | 3.25 |
| $\sigma_{G1}$ (km s$^{-1}$) | | 846 | 864 | 400 | 313 |
| $\epsilon_{G1}$ | $\cdots$ | 0.23 | 0.23 | 0 | 0 |
| $\theta_{G1}$ ($^\circ$) | $\cdots$ | –8 | –8 | $\cdots$ | $\cdots$ |
| $R_c$(G1) ($''$) | $\cdots$ | 0 | 0 | 0 | 0 |
| $\sigma_{C1}$ (km s$^{-1}$) | $\cdots$ | $\cdots$ | $\cdots$ | 2843 | 1000 |
| $\epsilon_{C1}$ | $\cdots$ | $\cdots$ | $\cdots$ | 0.61 | 0.55 |
| $\theta_{C1}$ (Deg.) | $\cdots$ | $\cdots$ | $\cdots$ | –33 | –5 |
| $R_c$(C1) ($''$) | $\cdots$ | $\cdots$ | $\cdots$ | 20.3 | 7.61 |
| $\Sigma_{C2}$ (g cm$^{-2}$) | 0 | 0 | 0 | 0 | 0.51 |

Fig. 1.— A deep color image of the Q2345+007 field, reconstructed from 22 hours of $B_j$ and R CCD exposures. The faintest galaxies visible on this print are 28 $B_j$ and 28 R magnitude. Sky rms noise is 31.2 and 30 mag arcsec$^{-2}$ in the two bands. North is up and east is left. This color image is 200″ across.

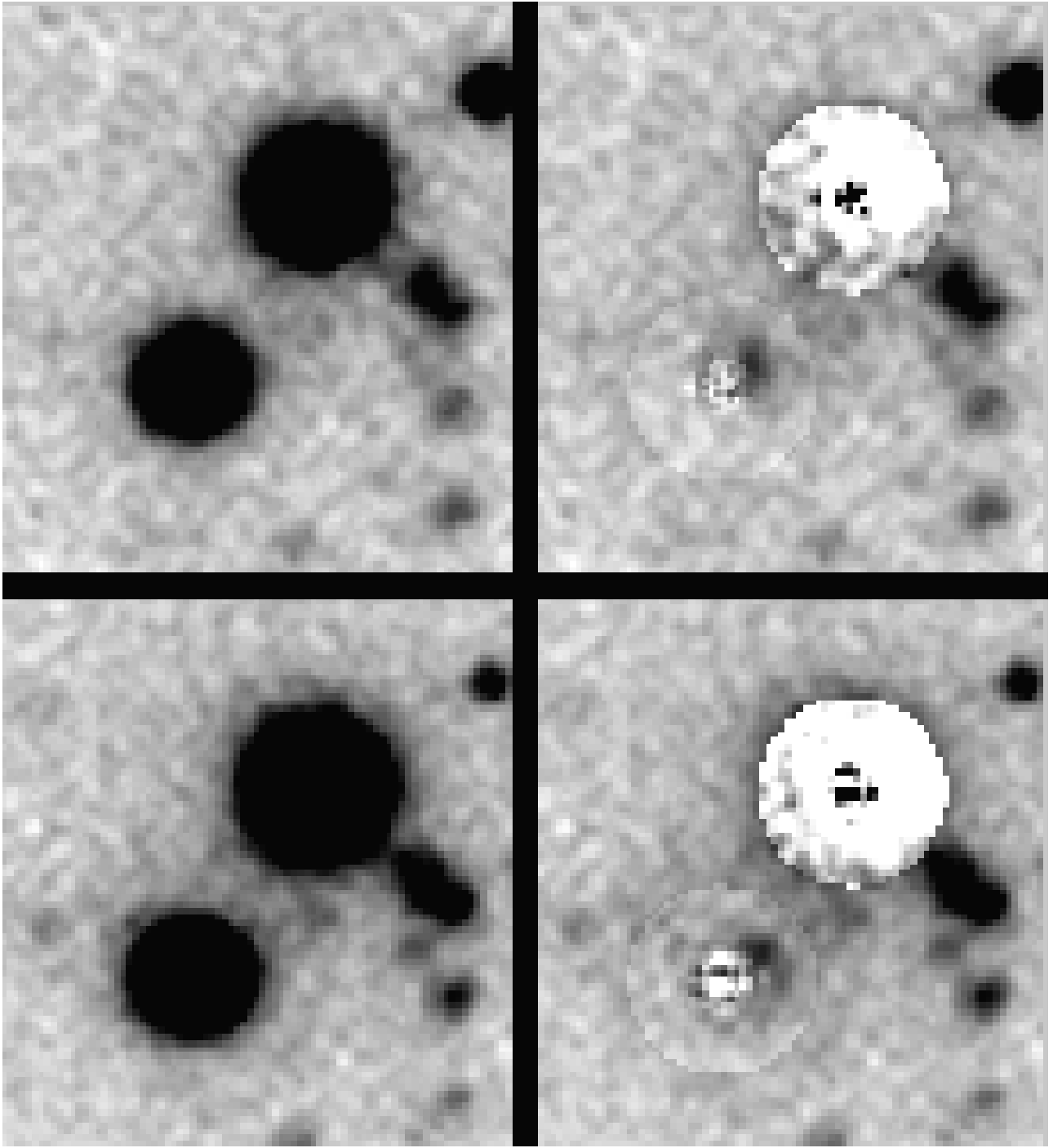

Fig. 2.— CCD image of the A and B images of the QSO (north is at the right and east is at the top, each panel is $16.2'' \times 18.2''$). The upper two panels are R band and the lower two panels are $B_j$ band. The right side shows the QSO-subtracted images. In both the $B_j$ and R band images there is the clear presence of a faint source (G1) located $1.3'' - 1.5''$ from image B ($1.0'' - 1.2''$ north and $0.8''$ east). The center is located $0.4'' - 0.6''$ northwest of the line joining A to B.

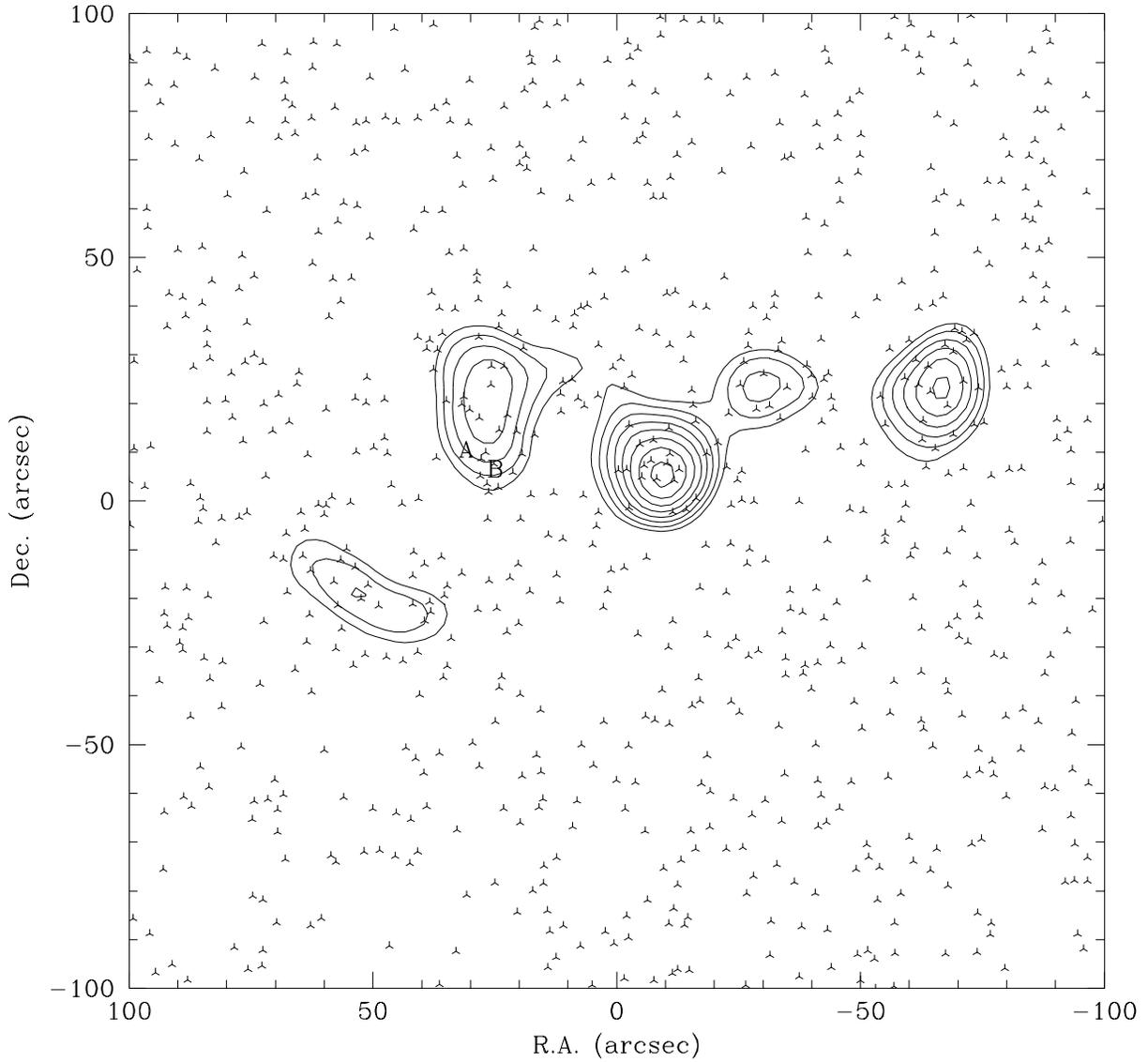

Fig. 3.— Contour diagram of number density for galaxies with $24 < B_j < 28$ and $B_j - R \leq 1.5$. North is at the top and east is to the left. The outermost contour corresponds to a surface density of 110 arcmin$^{-2}$ and adjacent contours differ by 5 arcmin$^{-2}$. The mean density for the region is 73 arcmin$^{-2}$. The points indicate galaxy positions and the quasar A and B images are marked.